\documentclass[prl,aps,showpacs,twocolumn,superscriptaddress]{revtex4}
\usepackage[utf8]{inputenc}
\usepackage[T1]{fontenc}                
\usepackage{amsmath, amsfonts, stmaryrd,dsfont,graphicx,physics,hyperref,bm}
\usepackage{subfigure}
\usepackage[capitalise]{cleveref}
\def\mb{\mathbf}
\def\a{\alpha}

\def\m{\mu}

\def\s{\sigma} 
\def\6{\partial} 
\def\d{\delta}

\def\g{\gamma}

\def\e{\epsilon}

\def\p{\pi}
\def\q{\theta}

\newcommand{\ea}[1]{\begin{align}#1\end{align}}
\newcommand{\eq}[1]{\begin{equation}\begin{split}#1\end{split}\end{equation}}

\def\f{\varphi} 
\def\l{\lambda}

\def\e{\epsilon}

\def\uar{\uparrow}
\def\dar{\downarrow}

\begin{document}
\title{Spin-accumulation induced magnetic texture in a metal-insulator bilayer}
\author{Dion M.F. Hartmann}\email[E-mail adress: ]{d.m.f.hartmann@uu.nl}\affiliation{Institute for Theoretical Physics, Utrecht University, Leuvenlaan 4, NL-3584 CE Utrecht, The Netherlands}
\author {Andreas Rückriegel}\affiliation{Institute for Theoretical Physics, Utrecht University, Leuvenlaan 4, NL-3584 CE Utrecht, The Netherlands}
\author {Rembert A. Duine}\affiliation{Institute for Theoretical Physics, Utrecht University, Leuvenlaan 4, NL-3584 CE Utrecht, The Netherlands}\affiliation{Department of Applied Physics, Eindhoven University of Technology, P.O. Box 513, 5600 MB Eindhoven, The Netherlands}
\begin{abstract}
We consider the influence of a spin accumulation in a normal metal on the magnetic statics and dynamics in an adjacent magnetic insulator. In particular, we focus on arbitary angles between the spin accumulation and the easy-axis of the magnetic insulator. Based on Landau-Lifshitz-Gilbert phenomenology supplemented with magnetoelectronic circuit theory, we find that the magnetic texture twists into a stable configuration that turns out to be described by a virtual, or image, domain wall configuration, i.e., a domain wall outside the ferromagnet. We show that even when the spin accumulation is perpendicular to the anisotropy axis, the magnetic texture develops a component parallel to the spin accumulation for sufficiently large spin bias. The emergence of this parallel component gives rise to threshold behavior in the spin Hall magnetoresistance and nonlocal magnon transport. This threshold can be used to design novel spintronic and magnonic devices that can be operated without external magnetic fields.
\end{abstract}
\maketitle
\textit{Introduction.} ---
The use of propagating spin waves, or magnons, to transmit and process information has the potential advantage of lower energy consumption over electronic currents. Especially insulating ferromagnets (IFM), such as yttrium-iron garnet (YIG), are able to accommodate a spin current efficiently as the damping of the magnetic dynamics is relatively low \cite{Thi18}. This has raised an increased interest in the possibilities of magnonic devices and how these could replace current electronic devices \cite{Cor15,Wu16}. Specifically, the behavior of magnons in magnetic domain wall textures can have promising applications \cite{San15,Wag16}. 

A typical experiment achieves transfer of angular momentum into an IFM through a spin current from a normal metal (NM) lead, usually platinum, by generating a spin accumulation at the interface by the spin-Hall effect \cite{Thi18,Hir99,Cor16b,Hua12}. The angle of this spin accumulation with respect to the magnetization at the interface determines the efficiency of spin current injection. In this paper we consider the effect of a sufficiently large spin bias which locally affects the magnetic texture and thereby the transfer of angular momentum. We propose an analytical solution for the magnetization texture of the IFM for a general orientation of the spin accumulation. 
Results for nonlocal magnon transport \cite{Zhe17} and the spin Hall magnetoresistance \cite{Che13,Dya71a} are derived. We find threshold behavior in both local and nonlocal setups for a critical magnitude of the spin accumulation. This threshold behavior may be employed as a useful functionality in novel spintronic and magnonic devices that, as a result, do not require a cumbersome external magnetic field to acces their different states. While threshold behavior is commonly associated with spin superfluidity \cite{Han18,Tak14}, our results show a threshold that is related to a change in the stable magnetic texture, and not to a spin superfluid state.


\begin{figure}
	\includegraphics[width=\linewidth]{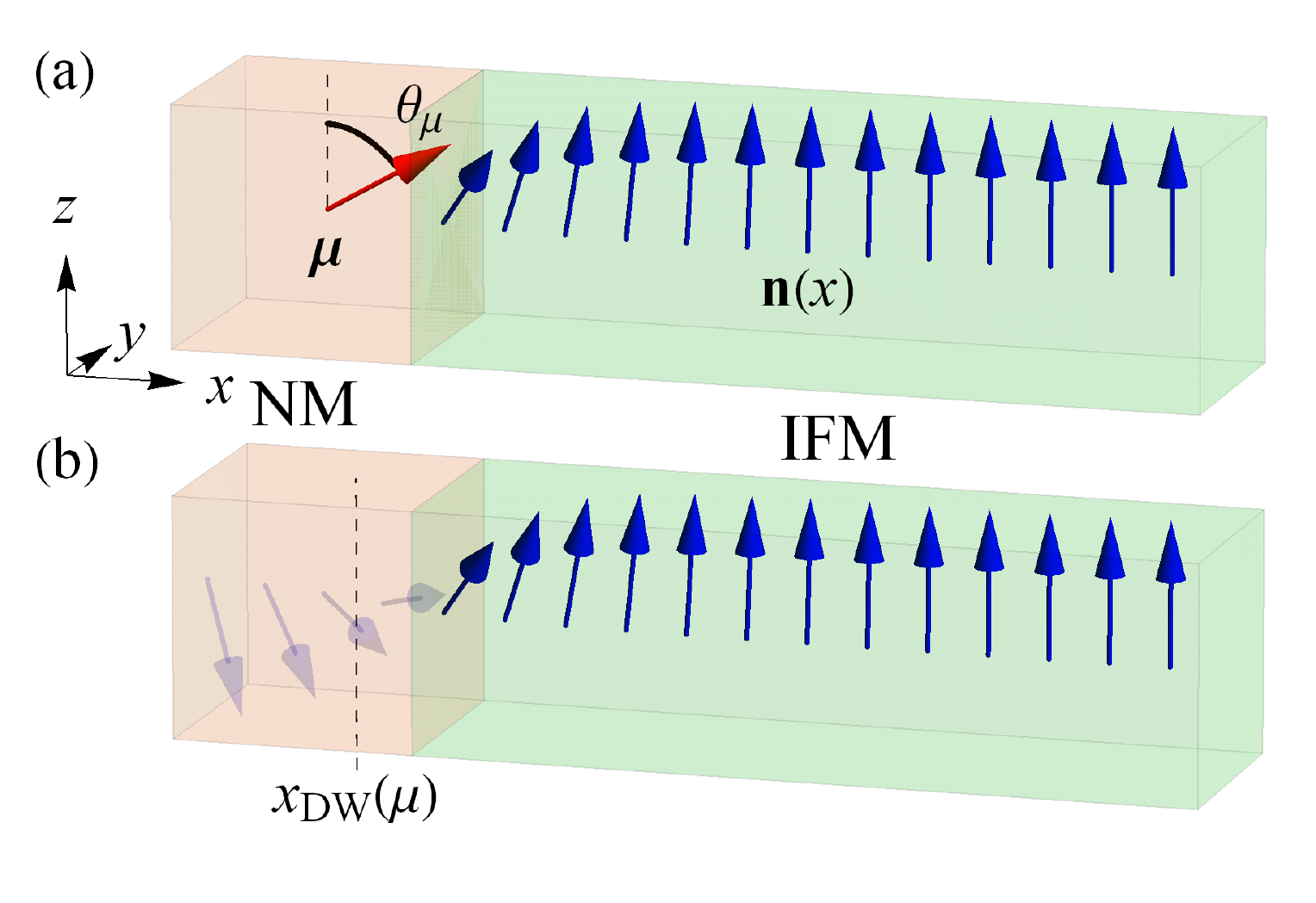}
	\caption{\label{fig:sys}
	(Color online) 
	(a): The magnetic texture $\mb{n}(x)$ (blue arrows) of the semi-infinite IFM nanowire (green region) with an easy-axis anisotropy in the $z$ direction. In the NM (orange region) an electric current generates a spin accumulation $\bm{\m}$ with polar angle $\q_\m$ at the interface (red arrow) that deforms the magnetic texture. 
	(b): The opaque arrows in the NM region are virtual and illustrate that the magnetic texture is that of two oppositely oriented domains with the center of the virtual domain wall, $x_{\textrm{DW}}$, outside the IFM. Such a virtual domain wall solution is found analytically for any magnitude and orientation of the spin accumulation.
	}
\end{figure}

\textit{Equations of motion.} ---
A one dimensional semi-infinite IFM nanowire with an interface with a nonmagnetic metal at $x=0$ is studied. At the interface a spin accumulation $\bm{\mu}$ is generated, e.g. by means of the spin-Hall effect, which results in a boundary condition on the spin current in the ferromagnet. A possible configuration of the system is illustrated in \cref{fig:sys} (a). Our aim is to determine the magnetic texture of the ferromagnet and its stability as a function of $\bm{\mu}$.
We define $\mb{n}=\mb{M}/M_{\mathrm{s}}$ as the unit vector in the direction of the magnetization, where $M_{\mathrm{s}}$ is the saturation magnetization. 
The energy of our system is given by
\eq{\label{eq:ene}
E=\int_V dV \frac{1}{2}\left(A|\6_x\mb{n}|^2-K n_z^2\right),
}
with $V$ the volume of the IFM, $A$ the spin stiffness,
 $K>0$ the easy-axis anisotropy 
and $n_z=\bm{\hat{z}}\cdot\mb{n}$. We consider an easy $z$ axis anisotropy, but the results apply to other easy-axis directions similarly. The Landau-Lifschitz-Gilbert (LLG) equation supplemented with spin-transfer torques and spin-puming terms that follows from magnetoelectronic circuit theory reads \cite{Gil04,Tse02}
\eq{\label{eq:LLG1}
	(1+\a_{\mathrm{G}}\mb{n}\times)\dot{\mb{n}}=&-\g\mb{n}\times\mb{H}_{\mathrm{eff}}\\
	&-\d(x)\frac{g^{\uar\dar}}{4\p s}\left[\mb{n}\times(\mb{n}\times\frac{\bm{\mu}}{\hbar}+\dot{\mb{n}})\right].
}
The left hand side describes the damped time evolution of $\mb{n}$, where $\a_{\mathrm{G}}$ is the dimensionless phenomenological Gilbert damping constant. The first term on the right hand side is the torque due to effective magnetic field
$\mb{H}_{\mathrm{eff}}$ which is given by 
\eq{
	\mb{H}_{\mathrm{eff}}=-\frac{1}{M_{\mathrm{s}}}\frac{\d E}{\d\mb{n}}=\frac{1}{M_{\mathrm{s}}}(A\6_x^2\mb{n}+K n_z \bm{\hat{z}}),
}
where $\g>0$ is the gyromagnetic ratio.
The second is the interfacial spin transfer torque and spin pumping respectively, where $g^{\uar\dar}$ is the interface spin flip scattering per surface area, i.e., the spin-mixing conductance, and $s$ the spin density.

The characteristic length scale of the ferromagnet is the exchange length $\l=\sqrt{A/K}$ and the ferromagnetic resonance frequency $\omega_{\textrm{F}}=\g K/M_{\mathrm{s}}$ sets the timescale.
Finally, we define $\a(x)=\a_{\mathrm{G}}+\l\d(x)\a'$, with $\a'=\frac{g^{\uar\dar}}{4\p\l s}$, so that the LLG equation is written as
\eq{\label{eq:LLGdl}
	(1+\a\mb{n}\times)\6_{t}\mb{n}=-\frac{\g}{M_\mathrm{s}}\mb{n}\times\Big[&A\6_x^2\mb{n}+K n_z \bm{\hat{z}}\\
	&+\d(x)\l\a'\left(\mb{n}\times\frac{{\bm{\mu}}}{\hbar\omega_\textrm{F}}\right)\Big].
}
We integrate the LLG equation around an infinitesimal interval around the interface to obtain the boundary condition on the spin current density:
\eq{
	\mb{j}^s|_{x=0}&= \left.\frac{\hbar s}{2}\frac{\g A}{M_{\mathrm{s}}}\mb{n}\times\6_{x} \mb{n}\right|_{x=0}\\
	&=-\l\a'\frac{\hbar s}{2}\mb{n}\times\left.\left(\mb{n}\times\frac{\bm{\mu}}{\hbar\omega_\textrm{F}}+\dot{\mb{n}}\right)\right|_{x=0}.\label{eq:bcj}
}
Furthermore, we have the boundary condition $\mb{n}\rightarrow\hat{\bm{z}}$ as $x\rightarrow\infty$. Now we set out to obtain a solution to the bulk part of \cref{eq:LLGdl} and use that to satisfy the boundary condition (\ref{eq:bcj}).

\textit{Virtual domain wall solution.} ---
It turns out that the stationary magnetization profile that obeys \cref{eq:LLGdl} and the boundary conditions is similar to a domain wall (DW) texture, but with the DW position outside of the ferromagnet: the DW is a stationary solution to the bulk part of the LLG equation, and the freedom of the DW position allows us to satisfy the boundary conditions. We refer to this situation as a virtual DW. Such a DW solution is written in spherical coordinates as
\eq{
	\mb{n}_{0}=\bm{\hat{x}}\sin\q\cos\f +\bm{\hat{y}}\sin\q\sin\f+\bm{\hat{z}}\cos\q,
	}
with $\f$ a constant azimuthal angle throughout the nanowire and $\q$ the polar angle given by
\eq{\label{eq:DWp}
	\q=2\arctan(e^{(x_{\textrm{DW}}-x)/\l}).
}
Here $x_{\textrm{DW}}$ is the position of the DW. 

\begin{figure*}[t]
	\centering
	\includegraphics[width=\linewidth]{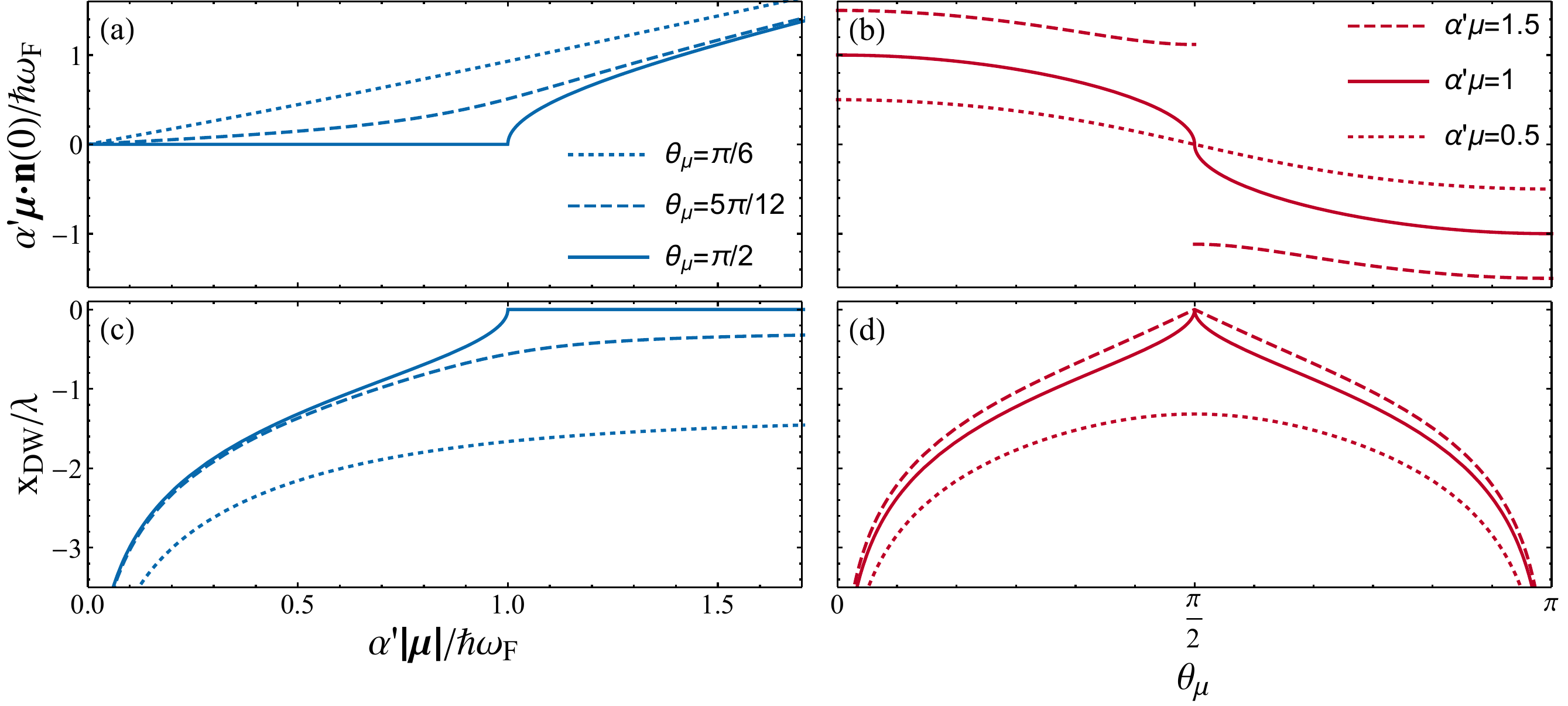}
	\caption{\label{fig:mut}
		(Color online)
		The component of the spin accumulation parallel to the magnetization at the interface ((a), (b)) and the virtual DW position $x_\textrm{DW}$ ((c), (d)) as a function of the value of the total spin accumulation $\m$ ((a), (c)) and its polar angle w.r.t. the $z$-axis ((b), (d)). These plots indicate the effect of the spin accumulation on the magnetic texture: As the spin accumulation increases, the virtual DW position approaches the interface, which will only be reached when $\q_\m=\p/2$, i.e., when $\bm{\m}$ is perpendicular to the anisotropy axis. For $|\bm{\m}|\leq\hbar\omega_{\textrm{F}}/\a'$, $\bm{\m}$ is also perpendicular the magnetization at the interface. But in the regime $|\bm{\m}|>\hbar\omega_{\textrm{F}}/\a'$ there will be a finite parallel component of the spin accumulation.
	}
\end{figure*}
Next, we study the boundary condition \cref{eq:bcj} of the spin current. For convenience we switch to a local spherical basis whose radial unit vector is given by $\mb{n}_{0}$. 
It follows that $\l\6_x\mb{n}_{0}=-\sin\q\bm{\hat\q}$. Hence,
\eq{
	\label{eq:jbc}
	\hspace{-5mm}\frac{2\mb{j}^s|_{x=0}}{ s \l\a'} =-\frac{\hbar\omega_\textrm{F}\sin\q_0}{\a'}\bm{\hat\f}=(\bm{\m}\cdot\bm{\hat\q})\bm{\hat\q}+(\bm{\m}\cdot\bm{\hat\f})\bm{\hat\f}|_{x=0}.
}
where $\q_0=\q(0)$. This gives us two equations:
\eq{
	\bm{\m}\cdot\bm{\hat\q}|_{x=0}=0\label{eq:mu1}\text{, and }
	\bm{\m}\cdot\bm{\hat\f}|_{x=0}=-\frac{\hbar\omega_\textrm{F}\sin\q_0}{\a'},
}
To solve these equations, we express $\bm{\m}$ in rescaled cylindrical coordinates:
$\m_z=\bm{\m}\cdot\bm{\hat{z}}/\hbar\omega_{\textrm{F}}$; $\m_R=\sqrt{(\bm{\m}\cdot\bm{\hat{x}})^2+(\bm{\m}\cdot\bm{\hat{y}})^2}/\hbar\omega_{\textrm{F}}$;
$\f_\m=\arctan{(\bm{\m}\cdot\bm{\hat{y}})/(\bm{\m}\cdot\bm{\hat{x}})}$.
Then we write
\ea{
	\bm{\m}\cdot\bm{\hat\q}&=\m_R\cos(\f-\f_\m)\cos\q-\m_z\sin\q;\label{eq:muq}\\
	\bm{\m}\cdot\bm{\hat\f}&=-\m_R\sin(\f-\f_\m)\label{eq:muf}.
}
From \cref{eq:muf}, we obtain an expression for the azimuthal angle $\f$ of the virtual DW in terms of $\f_\m$ and the polar angle $\q_0$ of the virtual DW at the interface:
\eq{
	\f-\f_\m=\begin{cases}
		\arcsin\left(\frac{\sin\q_0}{\a'\m_R}\right),& \text{ for }\m_z\geq0;\\
		\p-\arcsin\left(\frac{\sin\q_0}{\a'\m_R}\right),& \text{ otherwise. }
	\end{cases} \label{eq:fdw}
}
Note that $\f$ is only properly defined when $\m_R\neq0$. Indeed, if $\mu_R=0$ the boundary conditions fix $\sin\q=0$, 
i.e., the magnetization is homogeneous along the $z$ direction and an azimuthal angle is ill-defined. By inserting \cref{eq:fdw} into \cref{eq:muq}, we rewrite \cref{eq:mu1} and take the square to obtain
$
	(\a'^2\m_R^2-u)(1-u)=\a'^2\m_z^2 u$,
with $ u=\sin^2\q_0$. This is solved for $0\geq u\geq 1$ to obtain the expression for $x_{\textrm{DW}}$:
\eq{\label{eq:xdw}
	x_{\textrm{DW}}=-\text{arcsech}\left(\sqrt{\frac{1+\a'^2\m^2-\sqrt{(1-\a'^2\m^2)^2+4\a'^2\m_z^2}}{2}}\right),
}
where $\m=|\bm{\m}|/\hbar\omega_{\textrm{F}}$. Note that although the semi-infinite ferromagnet lies on the $x\geq 0$ axis, a virtual DW texture, i.e., $x_{\textrm{DW}}\leq0$, is the only physical solution, as this will minimize the energy of the system. This is seen directly from \cref{eq:ene} as the gradient in the first term is maximal around the virtual DW position. The role of $x_{\textrm{DW}}$ is merely to configure the virtual DW profile in such a way that the boundary conditions are met. The behavior of the magnetic texture as a function of $\bm{\m}$ is plotted in \cref{fig:mut}. The figure demonstrates the effect of the spin bias on the magnetic texture in terms of the virtual DW position and the component of the spin accumulation that is parallel to the magnetization at the interface.

A remarkable feature is that for increasing $|\bm{\m}|$ the virtual DW position approaches the interface. Precisely when $\q_\m=\p/2$, the virtual DW position will reach the interface when $|\bm{\m}|=\hbar\omega_{\textrm{F}}/\a'$. When $|\bm{\m}|$ increases further, the virtual DW position remains at the interface, but the azimuthal angle of the virtual DW now starts changing to pull the magnetization more parallel to the spin accumulation, resulting in the threshold behavior in the parallel component $\m_{||}=\bm{\m}\cdot\mb{n}_{0}|_{x=0}$ of the spin accumulation.

\textit{Spin Hall magnetoresistance.} ---
When applying an electric current through a NM|IFM system, the electrical resistance depends on the orientation of the magnetization of the IFM with respect to the current direction. The electric current $\mb{j}^e$ will generate a spin current $\mb{j}^s_x$ through the interface by the spin Hall effect. The magnitude of this current depends on the relative orientation of the magnetization of the IFM to the spin accumulation $\bm{\m}$ at the interface \cite{Dya71a,Che13}: The spin current is maximized (minimized) when the spin accumulation and magnetization at the interface are perpendicular (parallel) as then the most (no) angular momentum is transferred. As a result the resistivity in the NM is maximal (minimal) due to the inverse spin Hall effect. 

Considering \cref{fig:mut} (a), we expect a threshold effect in this spin Hall magnetoresistance of the normal metal when the angle $\q_j$ between the electrical current trough the NM and the anisotropy axis vanishes.
The applied electric field thus has a threshold value $E_{c}$, such that $|\bm{\m}|>\hbar\omega_\textrm{F}/\a'$, where the spin accumulation deforms the magnetic texture such that the transfer of angular momentum is reduced. 

Following \cite{Dya71a} we solve the coupled charge and spin current drift-diffusion equations as a function of the angle $\q_j$ between the electrical current and the anisotropy axis by inserting the boundary conditions for the spin current from \cref{eq:jbc}, assuming that $\bm{\m}$ obeys a diffusion equation (see \cref{app:smr}). In the large thickness (along the $x$ direction) limit for the NM and parallel current $\q_j=0$, the critical electric field for which the magnetic texture develops a component parallel to the spin accumulation, i.e., $|\bm{\m}|=\hbar\omega_\textrm{F}/\a'$, is given by
\eq{
\label{eq:Ecr}
E_c=\frac{\l s\g K}{\q_{SH}M_s}\left(\frac{2\p \hbar}{l_sg^{\uar\dar}e}+\frac{e}{\s}\right),
}
with $\q_{SH}$ the spin-Hall angle of the NM, $l_s$ the spin diffusion length, $e>0$ the elementary charge and $\s$ the electrical conductivity. To estimate this effect we consider a Pt|YIG interface where the critical electric field has a value of approximately $21$ V$/\m$m \cite{Thi18,Kli14,Sta09,Cor16}. 

\begin{figure}
	\includegraphics[width=\linewidth]{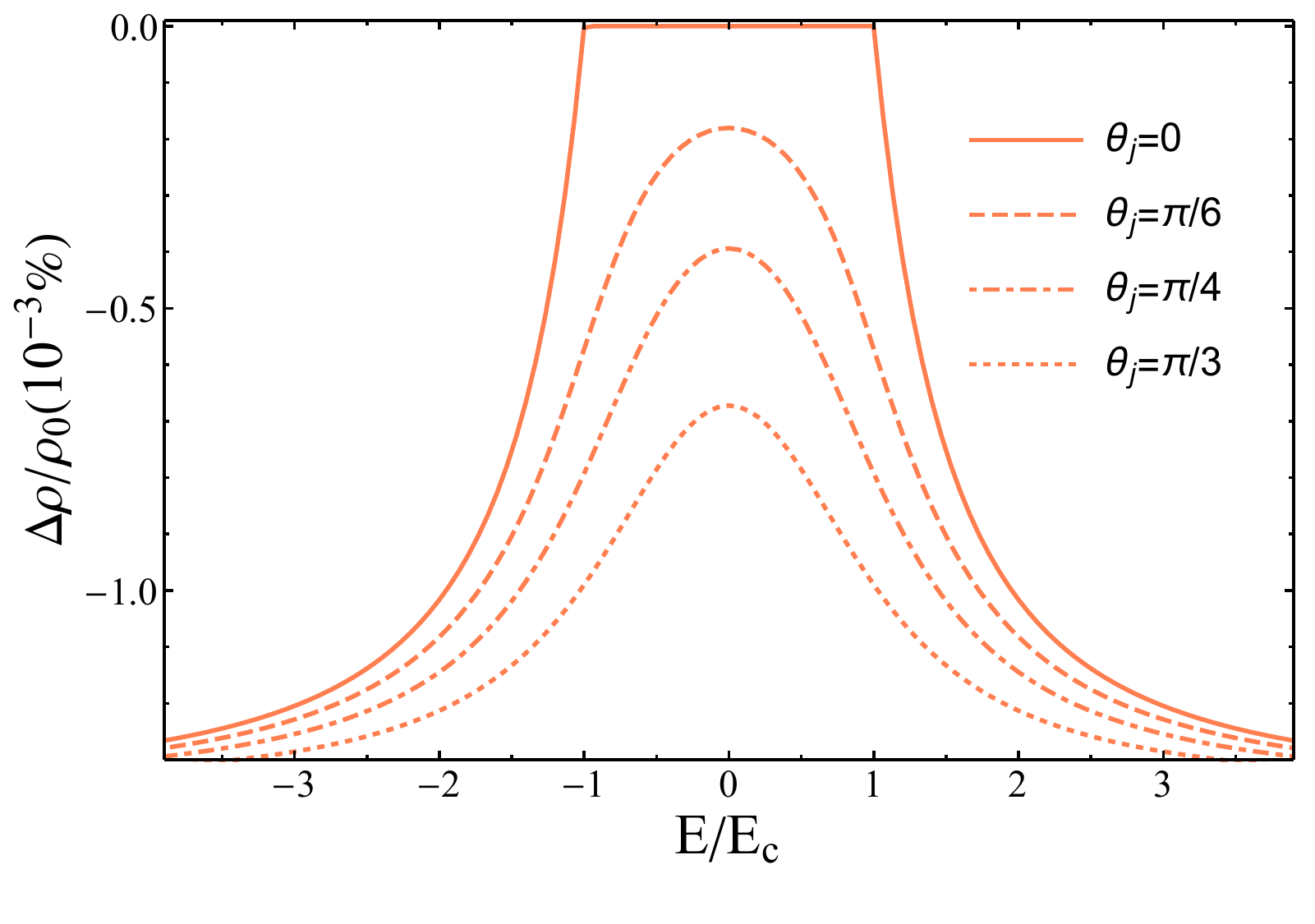}
	\caption{
		\label{fig:SMR}
		(Color online) Relative decrease in resistivity as a function of the normalized electric field $E/E_{c}$ for different angles $\q_j$ between the electric field and the anisotropy axis, where $E_{c}$ is the electric field that generates a spin accumulation $|\bm{\m}|=\hbar\omega_{\textrm{F}}/\a'$ at the interface. For $E\leq E_{c}$ and $\q_j=0$ the conductivity is not affected by the change in magnetization of the IFM as the magnetization remains perpendicular to $\bm{\m}$. For $E> E_{c}$ the magnetization at the interface aligns more with $\bm\m$ and the spin Hall magnetoresistance decreases.}
\end{figure}

In \cref{fig:SMR} we plot the normalized difference in resistance in the NM as a function of the applied electric field for a Pt|YIG interface. One clearly sees the threshold behavior of the resistance due to the change in magnetic texture as a function of the spin accumulation. 

\textit{Magnon transport.} ---
%
As we have seen, there is no transfer of spin when the spin accumulation and magnetization are parallel. Despite this, the IFM can accommodate the transfer of angular momentum by means of fluctuations (either thermal or quantum) in the form of spin waves, i.e., magnons. The magnons are injected and detected through spin-flip scattering at the interface with NM leads. The efficiency of the transfer of angular momentum is optimal when the spin accumulation is parallel to the magnetization at the interface. 
As a consequence, theshold behavior is expected in the nonlocal magnon transport signal. 
 
%
A typical experiment that quantifies the magnon transport attaches a lead at some position $x=d\gg\l$ and measures the electric current generated by the inverse spin-Hall effect \cite{Cor15}. 
To consider magnons, we add a perturbation to our stationary solution:
\eq{\label{eq:npe}
	\mb{n}=\mb{n}_{0}+\mb{\d n},\text{ with } \mb{\d n}=\bm{\hat{\q}}\d n_\q+\bm{\hat{\f}}\d n_\f,
}
where we make the anzats $|\d n_\q(x,t)|\ll1$ and $|\d n_\f(x,t)|\ll1$ are homogeneous along the $y$ and $z$ direction as we assume translation symmetry along the interface. 
The magnon field is defined as $\psi=\d n_\q+i\d n_\f$, and
thermal fluctuations are modeled by adding a stochastic field $\mb{h}$ to the LLG \cref{eq:LLGdl} \cite{Bro63,Zhe17}. Fourier transforming $\psi$ and $\mb{h}$, we obtain a Schrödinger-like equation from the linearized LLG equation:
\eq{
	(1+i\a)\omega \psi_\omega=\Big(&-\frac{\g A}{M_\mathrm{s}}\6_x^2+\omega_{\textrm{F}}\cos 2\q\\&+i\d(x)\l\a'\frac{\m_{||}}{\hbar}\Big)\psi_\omega
	- \g h_\omega.\label{eq:SEmw}
}
where $h=h_\q+ih_\f$. The second term on the right hand side plays the role of a local potential with a minimum at the virtual DW position.
The stochastic fields at the interfaces and in the bulk are combined into $\mb{h}=d\mb{h}^l\d(x)+\mb{h}^b+d\mb{h}^r\d(x-d)$, where each stochastic field obeys the fluctuation dissipation theorem \cite{Zhe17}: 
\ea{
	\langle h_\omega^\star(x) h_{\omega'}(x')\rangle&=4\p\frac{\a_{\mathrm{G}}\hbar\omega d}{\g M_\textrm{s}V} \frac{\d (x-x')\d (\omega-\omega')}{\tanh\left(\hbar\omega/2k_BT\right)};\\
	\langle h^{l \star}_{\omega} h^l_{\omega'}\rangle&=
	\frac{g^{\uar\dar}\hbar}{M_\textrm{s}^2 V d}  \frac{(\hbar\omega-\mu_{||})\d (\omega-\omega')}{\tanh\left(\frac{\hbar\omega-\m_{||}}{2k_BT}\right)};\\
	\langle h^{r \star}_{\omega} h^r_{\omega'}\rangle&=
	\frac{g^{\uar\dar}\hbar}{M_\textrm{s}^2 V d}\frac{\hbar\omega\d (\omega-\omega')}{\tanh\left(\hbar\omega/2k_BT\right)},
}
and
the temperature $T$ is assumed constant and equal in the bulk and at the leads as we are only interested in the nonlocal transport due to the spin bias. In this way, magnon dissipation at the boundaries and in the bulk is considered.

The observable we are interested in is the average spin current injected into the right lead at $x=d$. 
We have
\eq{
	\langle {j}^s \rangle =\left.
	\frac{M_\textrm{s}}{2\g}\Im\left\langle\psi^\star\left(\l\a'\dot{\psi}+d\g h^{r}\right)\right\rangle\right|_{x=d},	
}
where we defined $\langle {j}^s \rangle =
\langle \mb{j}^s \rangle \cdot\bm{n}_0|_{x=d}$. We use Green's functions to express $\psi$ in terms of the stochastic field and find an analytical solution using two types of solutions for the bulk part of \cref{eq:SEmw} \cite{San15,Zhe17}:
\eq{
	\label{eq:psi}
	\psi_{\omega,\pm}=(\mp i \l k(\omega)+\cos\q)e^{\pm ikx};\qquad
}
with $k(\omega)=\l^{-1}\sqrt{(1+i\a_{\mathrm{G}})\omega/\omega_{\textrm{F}}-1}$.
Remarkably, these magnon modes are stable regardless of the orientation and magnitude of the spin accumulation.
\begin{figure}[t]
	\includegraphics[width=\linewidth]{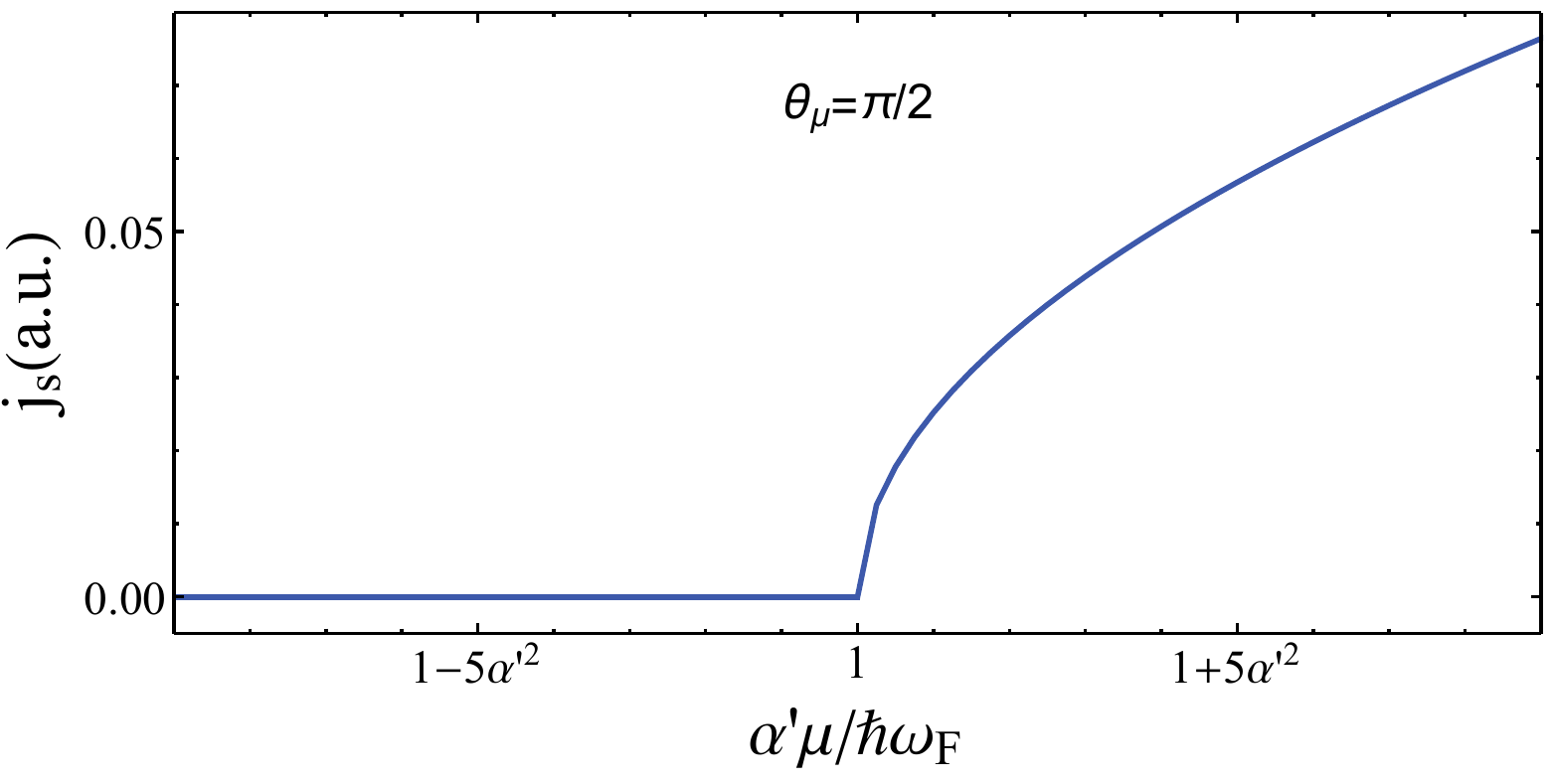}
	\caption{\label{fig:swp}
		(Color online)
		Threshold behavior in the spin current as a function of the bias. For $|\bm{\m}|>\hbar\omega_{\textrm{F}}/\a'$ there is a finite spin current when the spin accumulation is perpendicular to the anisotropy axis. This threshold behavior is caused by the deformation of the magnetic texture, generating a parallel component of the magnetization as is seen in \cref{fig:mut} (a). 
	}
\end{figure}

The result for the spin current at the right interface is written in the familiar Landauer-Büttiker form:
\eq{
	\hspace{-4mm}
	\langle {j}^s \rangle=
	\int\frac{d\omega}{2\p}\mathcal{T}(\omega)\left[N_B\left(\frac{\hbar\omega-\m_{||}}{k_BT}\right)-N_B\left(\frac{\hbar\omega}{k_BT}\right)\right],	
}
where $\mathcal{T}(\omega)= \a'^2\hbar d\omega(\omega-\m_{||}/\hbar)|G_\omega(0,d)|^2/V\omega_{\textrm{F}}^2$ is the transmission coefficient, $N_B(\e)=(e^\e-1)^{-1}$ is the Bose distribution function and $G_\omega(x,x')$ is the retarded magnon Green's function that solves \cref{eq:SEmw}.
Note that this spin current vanishes when $\m_{||}=0$. 

In \cref{fig:swp} the spin current injected into the right lead $x=d$ is plotted as a function of the spin accumulation at the left lead $x=0$, where the polar angle $\q_\m$ between the spin accumulation and the anisotropy axis is $\p/2$.
Our results show that for large bias, the spin accumulation affects the magnetic texture significantly. In particular, for $|\bm{\m}|>\hbar\omega_{\textrm{F}}/\a'$ there is a non-zero current even though the spin accumulation is perpendicular to the anisotropy axis. Such threshold behavior is also seen in experiments \cite{Han18}.

\textit{Conclusion.} ---
We have shown that a spin accumulation at the interface of a NM with an IFM affects the magnetic texture and thereby moderates the transfer of angular momentum across the interface. The magnetic texture is found analytically and is interpreted as a virtual DW, where the DW position always lies outside or at the boundary of the ferromagnet. 

Note that we do not fix the magnetization of the IFM at the interface as was done by Sitte \textit{et al.} \cite{Sit16} where a conducting ferromagnet is considered. There the authors demonstrate that by fixing the magnetization at the interface there is a critical current above which DWs are injected into the nanowire. Similarly, DWs are injected into our IFM system for sufficiently large spin accumulation and when the magnetization is fixed (see \cref{app:dwi}), which would physically correspond to a very large interface anisotropy.
%

Furthermore, we have shown that this interaction between the spin accumulation and the magnetization at the interface results in threshold behavior in spin Hall magnetoresistance and nonlocal magnon transport: When the spin accumulation exceeds the critical value $\hbar\omega_\textrm{F}/\a'$ the spin Hall magnetoresistance drops suddenly when the electric current is parallel to the anisotropy axis. Moreover, above the critical value a finite nonlocal magnon current can be measured even when the spin accumulation is oriented perpendicular to the anisotropy axis. These results provide a novel route to control both local and nonlocal spin transport signals via the electric current, without the need for an external magnetic field. We provide a possible geometry for an experimental setup in \cref{app:fat}.

We have assumed that the system size is large relative to the exchange length $\l$. For a smaller system, the exchange energy of the magnet cannot compensate the spin transfer torque, which leads to spin-torque oscillator instabilities \cite{Sla09} that prevent the formation of the virtual DW texture. Furthermore, we assume that the contact size of the biased lead is small compared to the distance between the leads to ensure that the magnons do not form a Bose-Einstein condensate when $\m_{||}>\hbar\omega_{\textrm{F}}/\a'$ \cite{Ben12}.

The electric field required to arrive at the threshold for a Pt|YIG bilayer is still two orders of magnitude higher than electric fields that have recently been applied in this kind of system \cite{Wim19}, but the expression for the critical electric field (\ref{eq:Ecr}) holds for any material, hence the threshold is more accessible for materials with a lower spin density, for example.

Remarkably, it is often argued that threshold behavior in nonlocal magnon transport indicates a metastable spin superfluid state \cite{Tak14,Han18,Son17}. However, we have demonstrated that even a stable magnetic texture may also lead to threshold behavior in the nonlocal magnon transport. We expect that an external magnetic field or a non-zero Dzyaloshinskii-Moriya interaction (DMI) might smoothen the threshold behavior as this will affect the azimuthal angle of the virtual DW. 

In future research our theory can be applied to interpret experimental results on such threshold behavior. Moreover, the model can be extended to antiferromagnets. Furthermore, the model can be enriched by considering the effects of a weak magnetic field or DMI.

\begin{acknowledgments}
	We acknowledge useful discussions with Julius Krebbekx and Geert Hoogeboom.
	R.D. is member of the D-ITP consortium, a program of the Dutch Organization for Scientific Research (NWO) that is funded by the Dutch Ministry of Education, Culture and Science (OCW). This work is funded by the European Research Council	(ERC). This work is part of the research programme of the Foundation for Fundamental Research on Matter (FOM), which is part of the Netherlands Organization for Scientific Research (NWO).
\end{acknowledgments}

\appendix
\setcounter{secnumdepth}{2}
	\section{Domain wall injection for insulating ferromagnets}
	\label{app:dwi}
	Recently it was shown by Sitte \textit{et al.} \cite{Sit16} that passing an electrical current trough a conducting ferromagnetic nanowire injects domain walls (DWs) into the magnet when the magnetization at the interface is fixed. Physically, the situation corresponds to an interface anisotropy that is so large that it dominates all other terms and fixes the magnetization direction at the interface. Here we show that the same result can be achieved for an insulating ferromagnet (IFM) by means of a spin accumulation at the interface with a normal metal. We will attempt to follow the derivation by Sitte \textit{et al.} as closely as possible.
	
	The system of interest is a semi-infinite ferromagnetic nanowire with an easy axis along the wire. At the left end of the wire ($x=0$) we fix the magnetization orientation $\mb{n}(0)=\hat{z}$. 
	The free energy of this system is given by
	\eq{
		F=\int_0^\infty\frac{1}{2}\left(A(\6_x\mb{n})^2+K\Pi(n_x)\right)dx
	}
	where $A$ is the exchange interaction, and $K$ the anisotropy for hte one dimenasional system (note that in the main text we consider a three dimensional system with translation invariance). $\Pi$ is a general form for the anisotropy, from which we only require $\Pi(0)=1$, $\Pi(1)=0$ and is monotonic and differentiable with $\Pi'(0)=0$.
	
	At $x=0$ there is a spin accumulation $\bm{\mu}=\mu_x \hat{x}+\m_y\hat{y}$ (an accumulation in the $z$-direction does not contribute). 
	The as in the main text, the LLG equation for this system reads
	\eq{
		(1+\a_G\mb{n}\times)\6_t\mb{n}=\g\mb{n}\times\mb{H}_{\mathrm{eff}}
		-\d(x)\frac{g^{\uar\dar}}{4\p s}\mb{n}\times\left(\mb{n}\times\frac{\bm{\mu}}{\hbar}+\dot{\mb{n}}\right).
	}
	Note that we set the gyromagnetic ratio $\g>0$ by convention, and use the same notation as in the main text.
	
	We aim to determine the critical spin accumulation energy below which there is a stable solution $\mb{n}$. Above this energy the dynamics will be slow and considered adiabatic, hence we will ignore dissipation. We thus reduce the LLG to 
	\eq{
		\6_t\mb{n}=-\frac{\g}{M_s}\mb{n}\times\frac{\d F_{\mathrm{eff}}}{\d\mb{n}},
	}
	where we defined
	\eq{
		F_{\mathrm{eff}}=F+\int_0^\infty\Omega_\mu dx,
	}
	with the Berry phase like term $\Omega_\mu$ satisfying
	\eq{
		\frac{\d}{\d\mb{n}}\left(\int_0^\infty\Omega_\mu dx\right)=\d(x)\frac{M_s}{\g\hbar}\frac{g^{\uar\dar}}{4\p s}(\mb{n}\times\bm{\mu}).
	}
	
	Now we consider $F_\mathrm{eff}$ as an action with corresponding Lagrangian
	\eq{
		\mathcal{L}=\frac{A}{2}(\6_x\mb{n})^2+\frac{K}{2}\Pi(n_x)+\Omega_\mu.
	}
	We may also define a Hamiltonian density
	\eq{
		\mathcal{H}=\frac{A}{2}(\6_x\mb{n})^2-\frac{K}{2}\Pi(n_x),
	}
	which should be conserved (w.r.t. $x$, i.e. translationally invariant). So evaluating at $x\rightarrow\infty$ we have that
	\eq{
		\label{eq:sit1}
		\frac{A}{2}(\6_x\mb{n})^2-\frac{K}{2}\Pi(n_x)=0.
	}
	
	Next, we consider the $x$ component of the LLG and integrate to obtain
	\eq{
		\int_0^\infty\left(\dot{n}_x+\frac{\g}{M_s}A(n_y\6_x^2n_z-n_z\6_x^2n_y)\right)dx=-\frac{g^{\uar\dar}}{4\p s\hbar}\m_x.
	}
	For a static solution $\dot{n}_x=0$, and with the boundary conditions $n_y(0)=n_z(\infty)=\6_xn_z(\infty)=0$ and $n_z(0)=1$ we use partial integration
	\eq{
		\int_0^\infty n_y\6_x^2n_z dx = \int_0^\infty n_z\6_x^2n_y dx + \6_x n_y|_{x=0},
	}
	to obtain
	\eq{
		\6_x n_y|_{x=0}=-\frac{g^{\uar\dar}}{4\p s}\frac{M_s\mu_x}{\g A\hbar}.
	}
	
	If we evaluate \cref{eq:sit1} at $x=0$, where we have $\6_x\mb{n} \perp \hat{z}$, and insert the above result, we obtain the condition 
	\eq{
		\mu_x\leq \frac{4\p s\g\hbar}{g^{\uar\dar}M_s}\sqrt{AK}=\frac{\hbar\omega_{\mathrm{F}}}{\a'},
	}
	for a stationary solution. For $\mu_x>\hbar\omega_{\mathrm{F}}/\a'$ the texture thus becomes unstable, and we have checked numerically that domain walls are then injected into the insulating ferromagnet. 
	
	\section{Properties of the domain wall profile}
	\label{app:dwp}
	As stated in the main article a stationary solution to the bulk part of the LLG equation is the domain wall, written in spherical coordinates as
	\eq{
		\mb{n}_{DW}=\bm{\hat{x}}\sin\q\cos\f +\bm{\hat{y}}\sin\q\sin\f+\bm{\hat{z}}\cos\q,
	}
	with $\f$ a constant azimuthal angle throughout the nanowire and $\q$ the polar angle given by
	\eq{\label{eq:DWp}
		\q=2\arctan\left(e^{(x_{DW}-x)/\l)}\right).
	}
	Here $x_{DW}$ is the position of the domain wall (DW) and $\l$ the DW width. To show that this indeed is a solution to the LLG equation, we first prove some usefull identities.
	For clarity we define $X=(x_{DW}-x)/\l$
	\begin{itemize}
		\item We have the following identities		
		\ea{
			\sin(\q)&=\sech(-X);\label{eq:id1}\\
			\cos(\q)&=\tanh(-X).\label{eq:id2}
		}
		Indeed $\sech^2(-X)+\tanh^2(-X)=1$, so there must be some angle $\q$ such that $\sin(\q)=\sech(-X)$ and $\cos(\q)=\tanh(-X)$. Now we will show that this is satisfied by \cref{eq:DWp}. This angle $\q$ must satisfy 
		\eq{\label{eq:i1}
			\tan(\frac{\pi}{2}-\q)=\cot(\q)=&\frac{\tanh(-X)}{\sech(-X)}\\
			=&\sinh(-X)=\frac{e^{-X}-e^{X}}{2}.
		}
		Now we can find $a$ and $b$ such that \eq{
			\tan(a)&=e^{-X}, \qquad \text{ and}\\ 
			\tan(b)&=e^{X}=\cot(a)=\tan(\frac{\pi}{2}-a).
		} 
		That is 
		\eq{
			a&=\frac{\pi}{2}-\arctan(e^{X}), \qquad \text{ and}\\ 
			b&=\arctan(e^{X}).
		}
		Now continuing \cref{eq:i1}
		\eq{
			\frac{e^{-X}-e^{X}}{2}&=\frac{\tan(a)-\tan(b)}{1+\tan(a)\tan(b)}\\
			&=\tan(a-b)\\
			&=\tan(\frac{\pi}{2}-2\arctan(e^{X})),
		}
		where we used the difference formula of the tangens for the last identity of the first line. Thus we obtain that indeed $\q=2\arctan(e^{X})$. $\square$
		\item The following identities hold
		\ea{\l\6_x\q&=-\sin\q;\label{eq:id3}\\
			\l\6_x\sin\q&=-\cos\q\sin\q;\label{eq:id4}\\
			\l^2\6^2_x\sin\q&=\cos 2\q\sin\q;\label{eq:id5}\\
			\l\6_x\cos\q&=\sin^2\q;\label{eq:id6}\\
			\l^2\6_x^2\cos\q&=-2\sin^2\q\cos\q;\label{eq:id7}\\
			&=(\cos 2\q-1)\cos\q.\nonumber
		}
		We only need to show the first identity and the rest will follow trivially. Note that from \cref{eq:id1,eq:id2} we have $e^{X}=\tan\q/2$, so
		\ea{
			\l\6_x\tan\q/2&=-e^{X}=-\tan\q/2\\
			&=\l\frac{\6_x\q}{2}\cos^{-2}\q/2.
		}
		Hence, rewriting the above gives 
		\eq{
			\l\6_x\q=-2\sin\q/2\cos\q/2=-\sin\q. 
		}
	\end{itemize}
	
	\noindent
	To show that $\mb{n}_{DW}$ indeed satisfies the LLG equation we derive
	\eq{
		\l^2\6_x^2\mb{n}_{DW}+(\mb{n}_{DW}\cdot\bm{\hat{z}})\bm{\hat{z}}=\mb{n}_{DW}\cos 2\q,
	}
	which is obtained readily with the help of \cref{eq:id5,eq:id7}. As this is clearly parallel to $\mb{n}_{DW}$, all terms in the LLG equation vanish. 

	\section{Spin Hall magnetoresistance}
	\label{app:smr}
	In this section we compute the resistance of a normal metal wire connected to a ferromagnetic insulator whose magnetization is parallel to the applied electric current. As the electric current will generate a spin accumulation at the interface, perpendicular to the magnetization, we expect an increase in the resistance once the applied voltage is above a threshold value, such that $|\bm{\m}|>\hbar\omega_{\textrm{F}}/\a'$ where the spin accumulation deforms the magnetic texture to allow for a nonzero spin current out of the normal metal.
	
	Considering the spin current $\mb{j}^s_x$ flowing perpendicular to the interface (the vector part describes the orientation of the spin), the relevant spin diffusion equations are \cite{Dya71a}:
	\ea{
		\label{eq:SHE1}
		j^e_z&=\frac{\s}{e}\6_z\m_e-\frac{\s\q_{SH}}{2e}\6_x\m_y;\\
		\label{eq:SHE2}
		\frac{\hbar}{2e}\mb{j}^s_{x}&=-\frac{\s}{2e}\6_x\bm{\m}-\frac{\s\q_{SH}}{e}\6_z\m_e\mb{\hat{y}}.
	}
	Here, $\sigma$ is the electric conductivity of the normal metal in units $\Omega^{-1}$m$^{-1}$, $e>0$ the elementary charge and $\q_{SH}$ the spin Hall angle. $\m_e$ is the electric potential, so $\6_z\m_e=F$ is the applied electric force. We rescale to make these equations dimensionless, defining $\tilde x= x/\l$, $\tilde t= t\omega_{\textrm{F}}$, $\tilde{\m}_e=\m_e/\hbar\omega_{\mathrm{F}}$ and $\tilde{F}=F\l/\hbar\omega_{\mathrm{F}}$. The currents will be normalized as $\tilde{\mb{j}}^e=\mb{j}^e/j^e_c$, with $j^e_c=\hbar\omega_{\mathrm{F}}\s/\l e$, and $\tilde{\mb{j}}^s=\mb{j}^s/j^{s}_c$, with $j^{s}_c=\hbar\omega_{\mathrm{F}} s \l/2$. Furthermore, we introduce the dimensionless constant $c_j=2ej^{s}_c/\hbar j^e_c$ such that, omitting the tildes for clarity, \cref{eq:SHE1,eq:SHE2} reduce to
	\ea{
		\label{eq:SHE1dl}
		j^e_z&=F-\frac{\q_{SH}}{2}\6_x\m_y;\\
		\label{eq:SHE2dl}
		c_j\mb{j}^s_{x}&=-\frac{1}{2}\6_x\bm{\m}-\q_{SH}F\mb{\hat{y}}.
	}
	The spin accumulation obeys the diffusion equation $\bm{\m}=\6_x^2\bm{\m}/l_s^2$, with $l_s$ the rescaled (w.r.t. $\l$) spin diffusion length. The solution is of the form
	\eq{
		\bm{\m}(x)=\mb{a}_-e^{-x/l_s}+\mb{a}_+e^{(x+t)/l_s},
	}
	with $t$ the rescaled thickness of the normal metal (along the $x$ axis). We have the following boundary conditions for \cref{eq:SHE2dl}:
	\ea{
		\mb{j}^s_{x}(x=-t)&=0;\\
		\mb{j}^s_{x}(x=0)&=\sin\q_0\bm{\hat{\f}},\label{eq:bc2}
	}
	where $\q_0=\q(x=0)$. The latter condition is obtained from the LLG equation in the ferromagnet, as discussed in the main text.
	For this particular orientation of the spin accumulation (that is $\m_z=0$), the solution for the DW position reduces to 
	\eq{
		x_{DW}=
		\begin{cases}
			0, & \text{ for } |\a'\m_y|\geq 1;\\
			-\textrm{arcsech}\sqrt{|\a'\m_y|}, & \text{ otherwise.}	
		\end{cases}
	}
	Using \cref{eq:id1} we obtain $\sin\q_0$ and thereby $\f=\f_\m+\arcsin(\sin\q_0/\a'\m_y)$. Thus the boundary conditions at $x=0$ become
	\eq{
		\mb{j}^s_{x}(x=0)=-\a'\bm{\m}
	}
	for $|\a'\m_y|\leq 1$. Otherwise
	\ea{
		j^s_{x,x}(x=0)&=-\frac{\sqrt{(\a'\m_R)^2-1}}{\a'\m_R}\approx-\frac{\sqrt{(\a'\m_y)^2-1}}{\a'\m_y};\\
		j^s_{x,y}(x=0)&=-\frac{1}{\a'\m_R}\approx-\frac{1}{\a'\m_y},
	}
	where we approximated around $\m_x=0$. We can solve the system exactly using numerics, or, with this approximation, obtain an analytical expression for $a_-$ and $a_+$, which allows us to determine the electrical current as a function of the applied electric force and study their non-linear dependence. 
	
	Furthermore, we generalize the result to the setup where the polar angle $\q_j$ between the electric current and the anisotropy axis is varied. Effectively, the boundary condition in \cref{eq:bc2} gets rotated over the angle $-\q_j$. The results are obtained numerically and discussed and illustrated in the main text.

	\section{Fluctuation assisted Transport}
	\label{app:fat}
	The observable we are interested in is the average spin current into the right lead at $x=d\gg\l$. We use $\d n_\q=(\psi+\psi^\star)/2$ and $\d n_\f=(\psi-\psi^\star)/2i$. Similar for the stochastic field $h^r_\q=(h^r+h^{r\star})/2$ and $h^r_\f=(h^r-h^{r\star})/2i$. When taking the average, terms linear in any of the independent components of the stochastic field $\mb{h}$ vanish. Note furthermore that $\psi$ does not depend on the radial component $\mb{n}_0\cdot\mb{h}$. Hence we find
	\ea{
		\langle\mb{\d n}\times\dot{\mb{\d n}}\rangle&=\bm{n}_0\Im\langle\psi^\star\dot{\psi}\rangle;\\
		\langle\mb{\d n}\times \bm{h}^r\rangle&=\bm{n}_0\Im\langle\psi^\star h^{r}\rangle.
	}
	The superscript indicates that we are considering the stochastic field at the right lead. Thus, averaging over the spin current at the right lead, we will determine
	\eq{
		\langle \bm{j}_s \rangle |_{x=d}\cdot\bm{n}_0=&\left.\frac{\hbar}{2}sd\left\langle\mb{\d n}\times\left(\frac{\l}{d}\a'\dot{\mb{\d n}}+\bm{h}^r\right)\right\rangle\right|_{x=d}\cdot\bm{n}_0\\
		=&\left.\frac{\hbar}{2}sd\Im\left\langle\psi^\star\left(\frac{\l}{d}\a'\dot{\psi}+h^{r})\right\rangle\right)\right|_{x=d},	
	}
	by working out to two terms on the right hand side separately. At the right interface we also have spin flip scattering, so now $\a(x)=\a_{\mathrm{G}}+\l\a'[\d(x)+\d(x-d)]$. Starting from the equation of motion given in the main text
	\eq{
		(1+i\a)\frac{\omega}{\omega_{\textrm{F}}} \psi_\omega=\Big(&-\l^2\6_x^2+\cos 2\q\\&+i\d(x)\l\a'\frac{\m_{||}}{\m_\mathrm{c}}\Big)\psi_\omega
		- \frac{h_\omega}{\omega_{\textrm{F}}}.\label{eq:SEmw2}
	}
	we will use the magnon's Greens function, defined by the equation
	\eq{\label{eq:eom}
		\left[(1+i\a)\frac{\omega}{\omega_{\textrm{F}}}+\l^2\6_x^2-\cos 2\q- i\d(x)\l\a'\frac{\m_{||}}{\m_\mathrm{c}}\right] G_\omega(x,x')\\
		=\l\d(x-x'),
	} 
	to express $\psi$ in terms of the stochastic field:
	\eq{\label{eq:gfp}
		\omega_{\textrm{F}}\psi_\omega(x)=-\int \frac{dx}{\l}'G_\omega(x,x') h_\omega(x'),
	}
	with $ h=d\d(x)h^l+h^b+d\d(x-d)h^r$, and
	\eq{\label{eq:ftp}
		\psi=\int\frac{d\omega}{2\p}\psi_\omega e^{i\omega t}.
	}
	Recall from the main text that the stochastic fields obey the fluctuation dissipation theorem \cite{Zhe17}, yielding
	\ea{
		\langle h_\omega^\star(x) h_{\omega'}(x')\rangle&=4\p\frac{\a_{\mathrm{G}}\hbar\omega d}{\g M_\textrm{s}V} \frac{\d (x-x')\d (\omega-\omega')}{\tanh\left(\hbar\omega/2k_BT\right)};\\
		\langle h^{l \star}_{\omega} h^l_{\omega'}\rangle&=
		\frac{g^{\uar\dar}\hbar}{M_\textrm{s}^2 V d}  \frac{(\hbar\omega-\mu_{||})\d (\omega-\omega')}{\tanh\left(\frac{\hbar\omega-\m_{||}}{2k_BT}\right)};\\
		\langle h^{r \star}_{\omega} h^r_{\omega'}\rangle&=
		\frac{g^{\uar\dar}\hbar}{M_\textrm{s}^2 V d}\frac{\hbar\omega\d (\omega-\omega')}{\tanh\left(\hbar\omega/2k_BT\right)},
	}
	
	For our purposes, we only need to consider the system of equations
	\ea{
		\left[(1+i\a_{\mathrm{G}})\frac{\omega}{\omega_{\textrm{F}}}+\l^2\6_x^2-\cos 2\q\right]G_\omega(x,d)&=0;\label{eq:SEG}\\
		\left[i\a'\left(\frac{\omega}{\omega_{\textrm{F}}}-\frac{\m_{||}}{\hbar\omega_{\textrm{F}}}\right)+\l\6_x\right]G_\omega(0,d)&=0;\label{eq:bcG1}\\
		\left[i\a'\frac{\omega}{\omega_{\textrm{F}}}-\l\6_x\right]G_\omega(d,d)&=1\label{eq:bcG2}.	
	}
	\begin{figure*}
		\includegraphics[width=\linewidth]{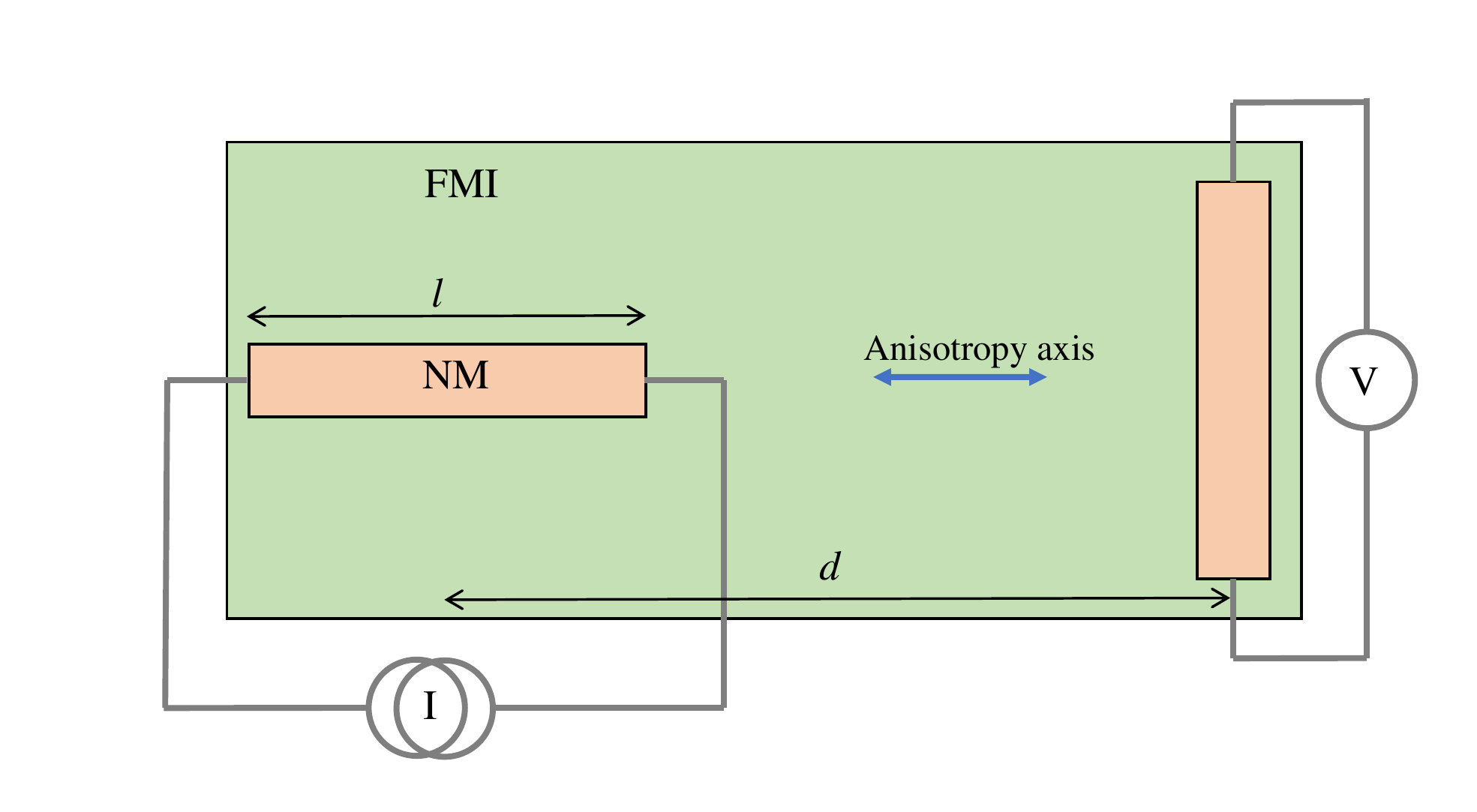}
		\caption{
			\label{fig:geo}
			(Color online) 
			A geometry for an experiment to measure the threshold behavior in the non-local magnon transport. The NM leads (orange) are attached perpendicular to each other in such a way that the injecting lead is parallel to the anisotropy axis (blue arrow) of the FMI (green). The theoretical model described in this paper is a good approximation if the distance between the leads $d$ is much larger than the length of the leads $l$.
		}
	\end{figure*}
	
	To find a solution for $G_\omega(x,d)$ we attempt 
	\eq{
		G_\omega(x,d)=&c_+(-i\l k+\cos\q)e^{ikx}\\
		&+c_-(i\l k+\cos\q)e^{-ik(x-d)},
	}
	with $k(\omega)=\l^{-1}\sqrt{(1+i\a_{\mathrm{G}})\omega/\omega_{\textrm{F}}-1}$ so that \cref{eq:SEG} is satisfied. The remaining two equation are used to determine $c_+$ and $c_-$ analytically. 
	
	We define the self energies 
	\ea{
		\Sigma_\omega(x,x')=&i\a_G\frac{\omega}{\omega_{\textrm{F}}}\frac{1}{\l}\d(x-x');\\
		\Sigma^l_\omega(x,x')=&i\a'\left(\frac{\omega}{\omega_{\textrm{F}}}-\frac{\mu_{||}}{\m_\mathrm{c}}\right)\d(x)\d(x-x');\\
		\Sigma^r_\omega(x,x')=&i\a'\frac{\omega}{\omega_{\textrm{F}}}\d(x-d)\d(x-x').
	}
	We can Fourier transform the time coordinate and express the wave function in terms of the Greens function and the total stochastic field to write
	\begin{widetext}
		\ea{
			\frac{\l}{d}\a'\langle\mb{\d n}\times\dot{\mb{\d n}}\rangle\cdot\bm{n}_0=&-2\int\frac{d\omega}{2\p}\int dx dx' dx'' dx''' \Sigma^r_\omega(x,x') G_\omega^\star(x',x'')\nonumber\\ &\qquad\left(\Sigma_\omega(x'',x''')F_\omega+\Sigma^l_\omega(x'',x''')F^l_\omega+\Sigma^r_\omega(x'',x''')F^r_\omega\right) G_\omega(x,x''');\\
			=&-2\int\frac{d\omega}{2\p}\Tr\left[\hat \Sigma^r_\omega\hat G_\omega^\dagger \left( \hat \Sigma_\omega F_\omega+\hat \Sigma^l_\omega F^l_\omega+\hat \Sigma^r_\omega F^r_\omega\right) \hat G_\omega\right].
		}
		Where we introduced a matrix (of infinite dimension) notation, where we interpret a function $f(x,x')$ as a matrix $\hat f$ with components labeled by $x$ and $x'$. The matrix product is defined by integration $(\hat f\cdot \hat g)_{xx'}=\int dx''f(x,x'')g(x'',x)$. The trace is defined by integrating over the diagonal components. Similarly, we express
		\ea{
			\langle\mb{\d n}\times \bm{h}^r\rangle\cdot\bm{n}_0=&2\int\frac{d\omega}{2\p}\Re\left(\Tr\left[\hat \Sigma^r_\omega\hat G^\dagger_\omega \left( \hat G_\omega\right)^{-1} \hat G_\omega\right]F^r_\omega\right);\\
			=&2\int\frac{d\omega}{2\p}\Re\left(\Tr\left[\hat \Sigma^r_\omega\hat G^\dagger_\omega \left(\hat \omega + \hat H + \hat \Sigma_\omega+\hat \Sigma^l_\omega +\hat \Sigma^r_\omega\right) \hat G_\omega\right]F^r_\omega\right);\\
			=&2\int\frac{d\omega}{2\p}\Tr\left[\hat \Sigma^r_\omega\hat G^\dagger_\omega \left( \hat \Sigma_\omega+\hat \Sigma^l_\omega +\hat \Sigma^r_\omega\right) \hat G_\omega\right]F^r_\omega.
		}
		Where we inserted an identity in the first line and read off the inverse matrix \cref{eq:eom},
		to be inserted in the second line, with $H(x,x')=\l^{-1}\d(x-x')(\l^2\6_x^2-\cos 2\q)$. 
		Note that the term proportional to $\hat\omega+\hat H$ drops out, as it is purely imaginary because $H$ is hermitian. Combining the two results we have
		\eq{
			\langle \mb{j}^s \rangle |_{x=L}\cdot\bm{n}_0&=
			-\hbar sd\int\frac{d\omega}{2\p}\Tr\left[\hat \Sigma^r_\omega\hat G_\omega^\dagger \left( \hat \Sigma_\omega (F_\omega-F_\omega^r)+\hat \Sigma^l_\omega (F^l_\omega-F_\omega^r)\right) \hat G_\omega\right];\\
			&=\hbar sd\int\frac{d\omega}{2\p}\a'\frac{\omega}{\omega_{\textrm{F}}}\left\{\a_{\mathrm{G}}\frac{\omega}{\omega_{\textrm{F}}}\int \frac{dx}{\l}|G_\omega(d,x)|^2\left[N_B\left(\frac{\omega}{T(x)}\right)-N_B\left(\frac{\omega}{T_r}\right)\right]\right.\\
			&\left.+\a'\left(\frac{\omega}{\omega_{\textrm{F}}}-\frac{\m_{||}}{\m_\mathrm{c}}\right)|G_\omega(d,0)|^2\left[N_B\left(\frac{\omega-\m_{||}}{T_l}\right)-N_B\left(\frac{\omega}{T_r}\right)\right]\right\},	
		}
	\end{widetext}
	Setting all temperatures equal yields the spin current as given in the main article.
	
	We propose a geometry for an experimental setup to measure the threshold behavior in the non-local magnon transport. A top view is illustrated in \cref{fig:geo}. The NM leads are attached perpendicular to each other and the magnetization anisotropy is parallel to the length of the injecting lead, so that the spin accumulation is perpendicular to the anisotropy axis for the injecting lead. At the detecting lead the magnetization will be perpendicular to the length of the lead.

	\begin{widetext}
		\section{Table of parameters}
		\label{app:tab}	
		\begin{table}[h!]
			\centering
			\begin{tabular}{l|l|r l|l}
				Constant & Symbol & Value & & Reference\\ 
				\hline
				Exchange stiffness & $A$ & $3.7\times10^{-12}$ & Jm$^{-1}$ & \cite{Kli14}\\
				Bulk Gilbert damping constant & $\a_G$ & $4.4\times10^{-4}$ & & \cite{Thi18}\\
				Interface Gilbert damping enhancement & $\a'=\frac{g^{\uar\dar}}{4\p\l s}$ & $5.8\times10^{-4}$ & & \\
				Spin mixing conductance & $g^{\uar\dar}$ & $3\times 10^{18}$ & m$^{-2}$ & \cite{Thi18}\\
				Gyromagnetic ratio & $\g$ & $1.79\times 10^{11}$ & s$^{-1}$T$^{-1}$ & \cite{Thi18}\\
				Spin-Hall angle & $\q_{SH}$ & 0.1 & & \\
				Perpendicular magnetic anisotropy & $K$ & $6.1\times 10^3$ & erg cm$^{-3}$ & \cite{Sta09}\\
				NM spin diffusion length & $l_s$ & $2$ & nm & \\
				Domain wall width & $\l=\sqrt{A/K}$ & $78$ & nm & \\
				Saturation magnetization & $M_s$ & $1.4\times10^5$ & Am$^{-1}$ & \cite{Kli14}\\
				Spin density & $s$ & $5.2\times10^{27}$ & m$^{-3}$ & \cite{Cor16}\\
				NM conductivity & $\s$ & $5\times10^6$ & Sm$^{-1}$ &	
			\end{tabular}
			\caption{\label{tab:par}
				Constants used to generate the numerical results.
			}
		\end{table}
	\end{widetext}

\bibliography{References_MagnonDW}
\end{document}